%% file: GroupDecoder0610.tex
\documentclass[12 pt, nodraft, onecolumn, journal]{IEEEtran}

\renewcommand{\baselinestretch}{1.5}

\usepackage[section]{placeins}
\usepackage{graphicx}
\usepackage{amsmath,amssymb, dsfont, epsfig}
\DeclareMathAlphabet\mathbfcal{OMS}{cmsy}{b}{n}
\usepackage{url}
\usepackage{color}
\usepackage{multirow}
\usepackage{amsmath}

\input{pream_bm}

\def\QED{\mbox{\rule[0pt]{1.5ex}{1.5ex}}}

\def\IR{\mathbb R}

\newcommand{\calS}{{\cal S}}

\newcommand{\C}{{\cal C}}

\newcommand{\D}{{\cal D}}

\newcommand{\M}{{\cal M}}

\begin{document}

\title{Hybrid Group Decoding for Scalable Video over MIMO-OFDM Downlink Systems}

\newpage

\author{Shuying Li \authorrefmark{1}\footnote{\authorrefmark{1} Shuying Li is with School of Electronics and Information Technology, Harbin Institute of Technology, Harbin, China 150001 (email: hitlishuying@gmail.com).}
\qquad Chen Gong \authorrefmark{2}\footnote{\authorrefmark{2} Chen Gong is with Qualcomm Research, San Diego, CA, USA 92121 (email: cg2474@caa.columbia.edu).}
\qquad Xiaodong Wang \authorrefmark{3}\footnote{\authorrefmark{3} Xiaodong Wang is with Electrical Engineering Department, Columbia University, New York, NY, USA 10027 (email: wangx@ee.columbia.edu).}
}

\maketitle
\allowdisplaybreaks

\begin{abstract}

We propose a scalable video broadcasting scheme over MIMO-OFDM systems.
The scalable video source layers are channel encoded and modulated into independent signal streams,
which are then transmitted from the allocated antennas in certain time-frequency blocks.
Each receiver employs the successive group decoder to decode the signal streams of interest by treating other signal streams as interference.
The transmitter performs adaptive coding and modulation, and transmission antenna and subcarrier allocation,
based on the rate feedback from the receivers.
We also propose a hybrid receiver that switches between the successive group decoder and the MMSE decoder depending on the rate.
Extensive simulations are provided to demonstrate the performance gain of the proposed group-decoding-based scalable video broadcasting scheme over the one based on the conventional MMSE decoding.
\renewcommand{\baselinestretch}{1.6}

\end{abstract}

\section{Introduction} \label{sec.Introduction}
High-quality video transmission over wireless channels has attracted extensive research interest as innovative communication techniques are being continuously developed.
Due to the high rate of the video sources, high spectrum efficiency transmission schemes are desired.
Due to the time-varying spectrum rate of the wireless fading channels,
the Scalable Video Codec (SVC) extension of H.264/AVC has been developed as a transmission-friendly video coding scheme~\cite{SVCOverview2},
where a video sequence is coded into several layers and proper layers are transmitted according to the current channel realizations.
Cross-layer wireless resource allocation for the SVC transmission has been addressed in a number of works,
including the joint source-channel coding (JSCC)~\cite{JointSourceandChannel1, JointSourceandChannel2}, the unequal error protection (UEP),
the content-aware video transmission~\cite{Content-aware2}, and resource allocation \cite{ResourceAllocation1} for video communications.
In the existing scalable video wireless communication system,
different coded video layers are encoded into different signals and transmitted in orthogonal channels~\cite{SVCMIMOModel, XuJunSVC}.

The MIMO-OFDM system with multi-stream multi-carrier transmission capability is a
key element in the current and near future standards,
such as 3GPP Long Term Evolution (LTE),
to achieve high peak throughput and spectral efficiency.
We consider an MIMO-OFDM system where each transmit antenna transmits independent signals.
The performance of the successive interference cancelation (SIC) decoding scheme in such systems has been studied in \cite{SICOFDMCluster2, SICOFDMOverview, SICOFDMThreshold}.
In this work, we further extend the SIC to the successive group decoder, where in each iteration the signal of one or more antennas is decoded,
until all desired signals are decoded.
Moreover, we perform rate allocation for the signal transmitted on each antenna.
It is known that for the SIC scheme, checking whether the decoding is correct and thus canceling only the correctly decoded signals
can provide further performance gain.
Therefore, in this work we employ the LDPC codes, which has error detection capability.

In \cite{PrasadOSGMAC, PrasadOSG, ChenGongCPGD, ChenGong&OmarCPGD},
the authors proposed the successive group decoder (SGD) for multiple-access and interference channels.
In this paper, we apply the SGD to the MIMO broadcast system, where the signal transmitted on each antenna is treated as a virtual user.
The SGD decodes the desired signal along with part of the interference in a successive manner from the received signal,
which is the superposition of the desired signal and the interference.

The contributions of this paper consist of the following.
We adopt the SGD for the MIMO broadcast system where each transmitting antenna transmits independent signals.
We propose a hybrid version of SGD that switches to MMSE decoding when the rate margin of the MMSE decoding exceeds a certain threshold.
Different from existing works, such as~\cite{ZanYang} that uses MMSE decoding, we adopt the hybrid decoding that outperforms MMSE decoding.
Different from most existing resource allocation works that allocates resources spanning over subcarriers and time slots including~\cite{ZanYang},
we consider a three-dimensional combination of the transmission resources, spanning over time slots, subcarriers, and transmission antennas.
We also propose the subcarrier and transmitting antenna allocation for the proposed SGD scheme.
Simulation results show significant peak signal-noise-ratio (PSNR) improvement of the reconstructed video, compared with the MMSE decoding.

The remainder of this paper is organized as follows.
In Section~\ref{sec.SystemDescription}, we describe the system model and background on the SVC communication system.
In Section~\ref{sec.HybridDecode}, we introduce the SGD and associated rate allocation for the MIMO system, as well as the hybrid group decoder.
In Section~\ref{sec.UEPandSubA}, we propose the resource allocation for the SGD.
Simulation results are given in Section~\ref{sec.SimulationResults}.
Finally, Section~\ref{sec.Conclusion} provides the concluding remarks.

\section{Background and System Descriptions}\label{sec.SystemDescription}

\subsection{Layered Video Broadcast over MIMO-OFDM System}\label{subsec.ChannelModel}

\begin{figure}[htbp]
\center{ \epsfxsize=5.5in \centerline{\epsffile{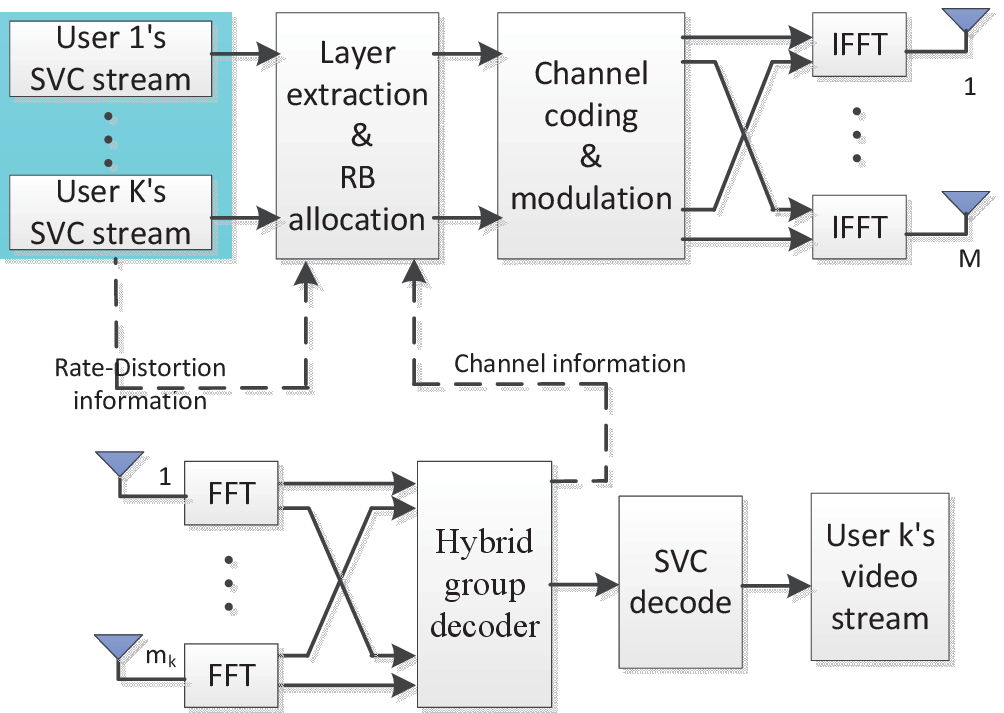}}
\caption{\small The MIMO-OFDM system for video broadcast.}
\label{Fig.Fig1MIMOdownlink}}
\end{figure}

Consider a single-cell $K$-user MIMO-OFDM broadcast system with $N$ subcarriers,
where an $M$-antenna base station broadcasts to $K$ users, and user $k$ is equipped with $m_k$ antennas.
Fig.~\ref{Fig.Fig1MIMOdownlink} depicts a MIMO-OFDM system for multiuser scalable video broadcast.
We consider a three-dimensional combination of the transmission resources, spanning over time, frequency, and transmit antennas.
The entire transmission consists of the following procedures.
\begin{itemize}
  \item Transmitter side: The base station obtains the SVC video data  of each user.
        Then, based on the achievable rates from the receivers' feedback and the rate-quality model,
        the transmitter performs the SVC video layer extraction and resource block (RB) allocation for the users,
        aiming to maximize the sum PSNR of the reconstructed video sequences.
        The channel coding and modulation are then applied to the extracted video source bits.
        The modulated symbols are then transmitted in the allocated resource blocks.
  \item Receiver side: Each receiver estimates  its achievable rate of the group decoder and feeds it back to the transmitter.
        The receivers also decode the video data using the SGD and then the
        SVC decoders reconstruct the video sequences.
\end{itemize}
We assume that the same channel code rate and modulation format is used for the data in an SVC layer,
even if the data is transmitted in different resource blocks and antennas.
We consider the LTE transmission scenario,
where a radio resource block (RB) spans over time slots and subcarriers.
Assume quasi-static block fading channels between the base station and the receivers,
where the channel gains are fixed during one transmission interval and change to another independent state afterwards.

\subsection{Scalable Video Coding (SVC)}\label{subsec.SVC}

Scalable Video Coding (SVC) is an extension to the H.264/AVC video codec,
which encodes a video sequence into a base layer and multiple enhancement layers with nested dependency structure.
The base layer provides a basic quality for the reconstructed video while the higher layers provide refined quality~\cite{SVCOverview2}.
A certain number of layers are transmitted according to the current channel condition, with more layers under better channel condition.

\begin{figure}[htbp]
\center{\epsfxsize=5.5in \centerline{\epsffile{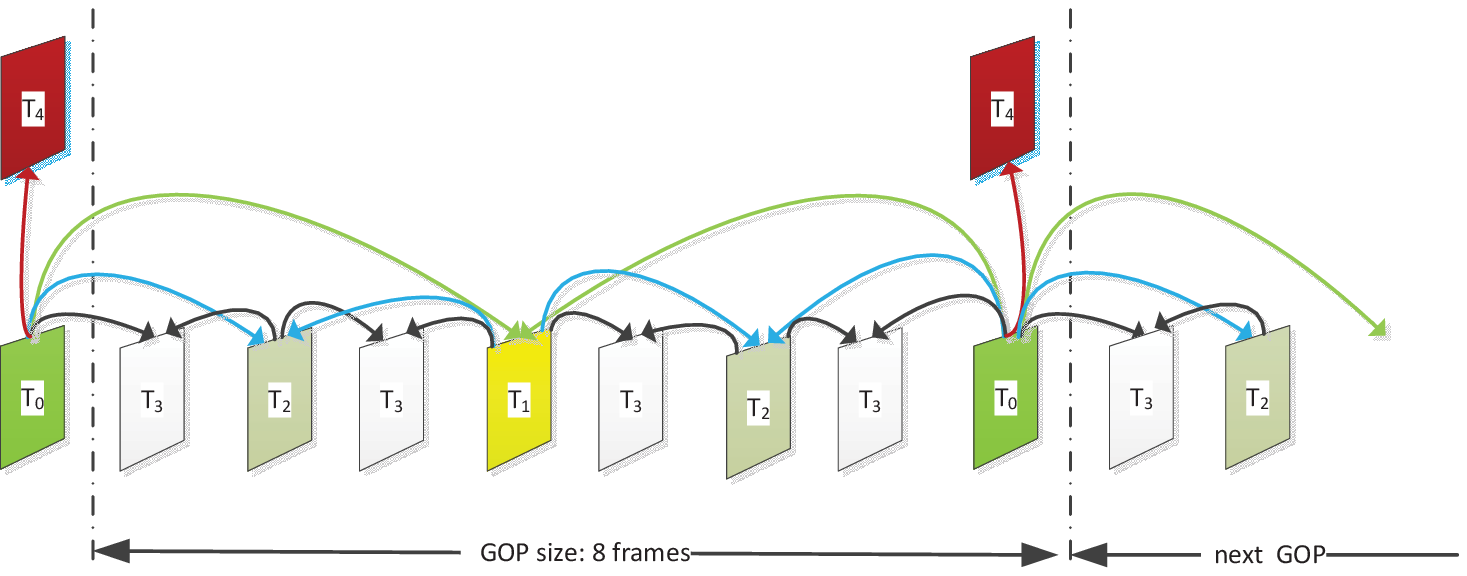}}
\caption{\small Illustration of layered video structure.}
\label{Fig.Fig2VideoLayers}}
\end{figure}

In this work, we assume that the video sequence is coded into several temporal layers and several enhanced quality layers.
We assume that a group of picture (GOP) consists of 8 frames, where the prediction structure is shown in Fig.~\ref{Fig.Fig2VideoLayers}.
Each frame is partitioned into $18$ slices, each macroblock row being a slice.

\subsection{End-to-End Video Distortion}

Let $f_n^p$ be the $p$-th pixel of the $n$-th frame in the original video sequence,
and $\tilde{f}_n^p$ be the corresponding reconstructed pixel at the decoder after error concealment.
The end-to-end mean squared distortion between the original video frame and the reconstructed video frame at the decoder is given by~\cite{E2EDistortion01705478},
\begin{equation}\label{equ.end2endDstortion}
    D_n = E\{[f_n^p - \tilde{f}_n^p]^2\}.
\end{equation}

The distortion $D_n$ is determined by many factors, including the quantization error in lossy video compression,
the substream extraction, the channel error, and the error concealment scheme employed at the decoder.
The accurate estimation not only requires the prior knowledge of the error concealment method, but also suffers from high computational complexity.
In this work, we use the Peak Signal to Noise Ratio (PSNR) (in dB) as the distortion metric for the reconstructed video.
The PSNR function of the reconstructed video in terms of the video rate $r$, denoted as $Q(r)$, can be approximated by \cite{ZanYang}
\begin{eqnarray}\label{equ.RQModel}
Q(r)=
\begin{cases}
    Q^0 + \beta^0(r-V_{0}), &r<V_{0}
    \cr Q^{l-1} + \beta^l(r-V_{l-1}), &V_{l-1}\leq r \leq V_{l}, l=1,\ldots,L
    \cr Q^L, &r \geq V_{L}
\end{cases}
\end{eqnarray}
where $\beta^l$ is the coefficient depending on the video sequence and codec setting;
$V_{l}$ is the total bit rate up to the $l$-th layer;
and $Q^l$ is the PSNR value up to the $l$-th layer ($l=0$ denotes the base layer).

We aim to maximize the average PSNR, $\bar{Q}= \frac{1}{K} \sum_{k=1}^{K}Q(r_k)$,
where $r_k$ is the rate allocation for receiver $k$, via the SVC layer extraction and RB allocation.
For convenience, the notations are listed in Table I.

\begin{table}[htbp]
\centering
\caption{\label{NotationTable}notation used in this paper}
\begin{tabular}{c||c}
\hline
Notation & Description\\
\hline
$Q(r)$ & PSNR with video rate $r$ \\
\hline
$V_{l}$ & The total bit rate up to $l$-th layer is extracted\\
\hline
 $Q^l$ & The PSNR value up to $l$-th layer is extracted \\
\hline
$\beta^l$ & PSNR-rate slope of the $l$-th layer\\
\hline
$K$ & Number of users\\
\hline
$N$ & Number of subcarriers\\
\hline
$M$ & Number of transmit antennas\\
\hline
$m_k$ & Number of the receive antennas of the $k$-th user\\
\hline
$\textbf{\emph{H}}_{i,k}$ & Channel matrix of the $i$-th RB of the $k$-th user \\
\hline
$\mu$ & The maximum group size of group decoding \\
\hline
$p_i$ & Number of decoding stages\\
\hline
$\mathcal{G}_{p_i}^i$ & The $p_i$-th decode order of receiver $i$\\
\hline
$\mathcal{\underline{G}}^i$ & Order partition of receiver $i$ \\
\hline
$P_{e}^l$ & Target BLER for the $l$-th layer\\
\hline
$r_{i,t,k}^l$ & rate on the $t$-th antenna of the $i$-th RB when the $l$-th layer of the $k$-th user is allocated\\
\hline
\end{tabular}
\end{table}

\section{Hybrid Group Decoding for MIMO Broadcast System} \label{sec.HybridDecode}

In this section, we propose a hybrid decoding scheme for MIMO broadcast systems.
We first define some notations.
Let $(\cdot)^\dagger$ denote the Hermitian transpose.
Let a calligraphic uppercase letter (e.g., $\mathcal{A}$) denote a finite set of integers.
Let an underlined calligraphic uppercase letter (e.g., $\mathcal{\underline{G}}$) denote the ordered partition of a set.

\subsection{Successive Group Decoder (SGD) for MIMO-OFDM-BC System}\label{subsec.SGDforVBS}

The SGD scheme was originally proposed for interference channels \cite{PrasadOSGMAC, PrasadOSG, ChenGongCPGD, ChenGong&OmarCPGD},
that exhibits significant performance gain over the conventional MMSE decoding.
For the MIMO broadcast system where each transmit antenna transmits an independent data stream, the received signal is the superposition of
the signals from all transmit antennas, which is of the same nature as that of the interference channel.
The SGD can be employed at each receiver where in each stage some layers of the video source are decoded while treating the undecoded layers as noise.
The decoded layers are then subtracted from the received signal, until all its desired layers are decoded.
In the remainder of this subsection, we provide an overview of the SGD for each RB, where the index of the RB is omitted.

Consider the discrete-time model of a slow-fading MIMO-OFDM broadcast system.
The base station is equipped with $M \geq 1$ antennas and broadcasting to $K$ users.
Each receiver $k$ is equipped with $m_k$ antennas.
Each transmission antenna transmits one OFDM symbol.
In particular,we consider an OFDM RB, where the received signal of the $k$-th receiver is given by,
\begin{equation}\label{eq.MIMOOFDMBCmodel}
    \by_{k} = \bH_{k}\bx + \bu_{k} = \sum_{t=1}^M \bh_{k}^t x^t + \bu_{k},
\end{equation}
where $\by_{k} = [\emph{y}_{k}^1, \emph{y}_{k}^2, \cdots, \emph{y}_{k}^{m_k}]^T$,
$\bx = [\emph{x}^1, \emph{x}^2, \cdots, \emph{x}^{M}]^T$,
and $\bu_{k} = [\emph{u}_{k}^1, \emph{u}_{k}^2, \cdots, \emph{u}_{k}^{m_k}]^T$ are the received signal, the transmitted signal and the AWGN, respectively,
and $\emph{\textbf{H}}_{k}$ denotes the $m_k \times M$ channel matrix.
We assume that $\mathbb{E}(|x^t|^2) = 1$ and incorporate the signal power into the channel realization vector $\bh_{k}^t$.
From (\ref{eq.MIMOOFDMBCmodel}), the MIMO broadcast channel can be treated as an equivalent interference channel,
where each transmission antenna is considered as a virtual transmitter.
We assume independent AWGN with the noise variance $\sigma^2$.
In the following we describe the SGD for decoding the information from each transmit antenna.

Let $\mathcal{M}= \{1,2, \cdots, M \}$.
For each receiver $k$, we say that a given ordered partition $\mathcal{\underline{G}}^k=\{\mathcal{G}_1^k,\cdots,\mathcal{G}_{p_{k+1}}^k \}$ of $\mathcal{M}$,
where $p_k$ is the number of decoding stages, is valid if the following three conditions are satisfied:
1) $|\mathcal{G}_m^k | \leq \mu$ for $m \in \{1, \cdots, p_k \}$, where $\mu$ is the maximum group size;
2) the rate vector $\mathcal{\bR}_{\mathcal{G}_m^k }$ is decodable at the $m^{th}$ stage of the successive decoding procedure for $m \in \{1, \cdots, p_k \}$;
3) the desired signal layers of receiver $k$ are decoded in the first $p_k$ stages, i.e., it belongs to $\bigcup^{p_k}_{m=1}\mathcal{G}_{m}^k$.

For a given valid partition $\mathcal{\underline{G}}^k$ of $\mathcal{M}$,
in the $m^{th}$ stage, the receiver jointly decodes the signals from antennas in $\mathcal{G}_m^k$ by treating $\{\mathcal{G}_{m+1}^k,\cdots,\mathcal{G}_{p_k+1}^k \}$ as additive noise and then subtracts the decoded messages in $\mathcal{G}_m^k$ from the received signal.
Note that, in the $m^{th}$ stage, we compute the noise covariance matrix
\be \label{eq.CovMat1}
\Sigma_{k,m} = \sigma^2\textbf{\emph{I}} + \sum_{q\in \cup^{p_k+1}_{e=m+1}\mathcal{G}_{e}^k} \emph{\textbf{h}}_{k}^q \emph{\textbf{h}}_{k}^{q \dagger},
\ee
and decode the information $\bx_{\mathcal{G}_m^k} \dff [x^q]_{q \in \mathcal{G}_m^k}$ from the following signal
\be \label{eq.CovMat2}
\br_{k,m} = \Sigma_{k,m}^{-1/2} \by^m_{k} = \Sigma_{k,m}^{-1/2}\emph{\textbf{H}}_{k,\mathcal{G}_m^k} \emph{\textbf{x}}_{\mathcal{G}_m^k} + \emph{\textbf{u}}_{k,m}, \ \bH_{k,{\cal G}^k_m} = [\bh^q_k]_{q \in {\cal G}^k_m}
\ee
where $\emph{\textbf{u}}_{k,m} \in \mathcal{N_C}(0,\textbf{\emph{I}})$ is the AWGN with unit variance,
and $\by^m_{k} = \emph{\textbf{H}}_{k,\mathcal{G}_m^k} \emph{\textbf{x}}_{\mathcal{G}_m^k}+\emph{\textbf{u}}_{k,\mathcal{G}_m^k}$ is the residue signal in the $m^{th}$ stage.

We define a rate outage as an event where in a decoding stage the rates of the signals to be decoded fall out of the corresponding achievable rate region.
Let $R_{t}$ be the transmission rate of the signal on transmitting antenna $t$ and $\bR \dff [R_{t}]_{1 \leq t \leq M}$.
We define the following rate margin for decoding $\mathcal{A}$ while treating $\mathcal{B}$ as noise for two disjoint subsets $\mathcal{A}$, $\mathcal{B} \subseteq \mathcal{M}$ as follows
\begin{equation} \label{eq.DeltaRate}
\varepsilon(\bH_{k},\mathcal{A},\mathcal{B},\bR)\triangleq \min_{\mathcal{D}\subseteq\mathcal{A},\mathcal{D}\neq\phi}\{ \frac {\vartriangle(\bH_{k},\mathcal{D},\mathcal{B},\bR)} {|\mathcal{D}|}\}, \quad \mathcal{A}\neq\phi ,
\end{equation}
with $\varepsilon(\emph{\bH}_{k},\phi,\mathcal{B},\bR)=0$ and
\begin{equation}
\vartriangle(\emph{\bH}_{k},\mathcal{D},\mathcal{B},\mathbf{\bR})\triangleq \log\left|\mathbf{\emph{I}}+ \emph{\bH}^{\dag}_{k,\mathcal{D}} \left(\mathbf{\emph{I}}+\emph{\bH}_{k,\mathcal{B}} \emph{\bH}^{\dag}_{k,\mathcal{B}}\right)^{-1}\emph{\bH}_{k,\mathcal{D}}\right|-\sum_{t\in\mathcal{D}}R_{t} .
\end{equation}
For the valid ordered partition $\mathcal{\underline{G}}^k=\{\mathcal{G}_1^k,\cdots,\mathcal{G}_{p_{k+1}}^k\}$, we define
\begin{equation}
\varepsilon(\emph{\bH}_{k},\mathcal{\underline{G}}^k,\bR)\triangleq\min_{1\leqslant m\leqslant p_k}\left\{\varepsilon\left(\emph{\bH}_{k},\mathcal{G}_m^k,\mathcal{M}\backslash\cup_{l=1}^m\mathcal{G}_l^k,\bR\right)\right\},
\end{equation}
as the minimum rate margin through the $p_k$-stage successive decoding.
The rate outage at receiver $k$ is equivalent to $\varepsilon(\emph{\textbf{H}}_{k},\mathcal{\underline{G}}^k,\bR)<0$.
Each receiver $k$ needs to find the optimal decoding order that maximizes the rate margin, i.e., finding
\begin{equation}\label{equ.OSGDorigenal}
    \varepsilon_{opt} = \max_{\mathcal{\underline{G}}^k} \varepsilon(\emph{\bH}_{k},\mathcal{\underline{G}}^k,\bR).
\end{equation}
 The SGD with such optimal decoding order is called the optimal SGD (OSGD).
 A greedy algorithm, Algorithm~1, can be used to solve \eqref{equ.OSGDorigenal}, which either declares an outage or identifies the optimal valid partition.
In each step, assuming the undecoded set to be $\mathcal{S}$, receiver $k$ finds the optimal set of the decoded user, denoted as $\mathcal{G}^*$, as follows
\begin{equation}\label{equ.OSGDorig2}
\mathcal{G}^* = \arg\max_{\mathcal{G}\subseteq \mathcal{S},\mathcal{G}\neq \phi} \varepsilon(\bH_{k},\mathcal{G}, \mathcal{S}\setminus\mathcal{G}, \bR).
\end{equation}
If in a step the selected $\mathcal{G}^*$ leads to the rate margin $\varepsilon(\emph{H},\mathcal{G^*}, \mathcal{S}\setminus\mathcal{G^*}, \bR)<0$,
then a rate outage event is declared.

\begin{minipage}[h]{6.5 in}
\rule{\linewidth}{0.3mm}\vspace{-.1in}
{\bf {\footnotesize Algorithm 1 - Greedy Partitioning for Fixed Rate $R$}}\vspace{-.2in}\\
\rule{\linewidth}{0.2mm}
{ {\small
\begin{tabular}{ll}
   \;1:& Initialize $\mathcal{S} = \mathcal{K}$, $\mathcal{G}_{opt}=\phi$ \\
   \;2:& Identify a group\\
   \;3:& $\mathcal{G}^* = \arg\max_{\mathcal{G}\subseteq \mathcal{S},|\mathcal{G}|\leq \mu_i, \mathcal{G}\neq \phi}\{\varepsilon(\bH_{k},\mathcal{G}, \mathcal{S}\setminus\mathcal{G}, \mathbf{R})\}$ \\
   \;4:& \textbf{If} $\varepsilon(\bH_{k},\mathcal{G^*}, \mathcal{S}\setminus\mathcal{G^*}, \mathbf{R})<0$, \textbf{then} \\
   \;5:& \quad declare a rate outage and stop;\\
   \;6:& \textbf{\bf Else}  \\
   \;7:& \quad update $\mathcal{S} \leftarrow \mathcal{S}\backslash \mathcal{G}^*$ and $\mathcal{G}_{opt} \leftarrow \{\mathcal{G}_{opt}, \mathcal{G}^*\}$\\
   \;8:& \quad \textbf{Until} $\mathcal{G}=\phi$ \\
   \;9:& \textbf{end if}
\end{tabular}}}\\
\rule{\linewidth}{0.3mm}
\end{minipage}\vspace{.2 in}\\

The optimal group search problem \eqref{equ.OSGDorigenal} can be solved using simple exhaustive search by enumerating all possible
nonempty set $\mathcal{G} \subseteq \mathcal{S}$ with $|\mathcal{G}| \leq \mu_j$.
Such an exhaustive method can be applied for small $\mu_j$, e.g., $\mu_j=1$ or $\mu_j=2$, which is the case for most practical scenarios.
Algorithm~2 can be efficiently applied to solve the optimal problem \eqref{equ.OSGDorigenal} for the large $\mu_j$ cases.

\begin{minipage}[h]{6.5 in}
\rule{\linewidth}{0.3mm}\vspace{-.1in}
{\bf {\footnotesize Algorithm 2 - Selecting an Optimal Group}}\vspace{-.2in}\\
\rule{\linewidth}{0.2mm}
{ {\small
\begin{tabular}{ll}
    \;1: & Initialize user set $\mathcal{S}$ and rates $\mathbf{R}_{\mathcal{S}}$\\
    \;2: & Let $\mathbfcal{S} \;\dff\;\{\mathcal{G}\subseteq\mathcal{S}:\mathcal{G}\neq\phi, |\mathcal{G}|=\mu_i$ or $\mathcal{G}=\mathcal{S}\}$ and set $\mathbfcal{S}_1 = \phi, \delta = -\infty.$\\
    \;3: & \textbf{For} each $\mathcal{G}\in \mathbfcal{S}$\\
    \;4: & \quad \textbf{\bf repeat}\\
    \;5: & \quad \quad Update $\mathbfcal{S}_1\longleftarrow\{\mathbfcal{S}_1,\mathcal{G}\}.$\\
    \;6: & \quad \quad Determine\\
     & \quad \quad \qquad$a=\min_{\mathcal{W}\subseteq\mathcal{G},\mathcal{W}\neq\phi}\Delta(\bH_{k},\mathcal{W},\mathcal{S}\backslash\mathcal{G},\mathbf{R}_{\mathcal{W}})$\\
     & \quad \quad and let $\hat{\mathcal{W}}$ be the set of the smallest cardinality\\
    \;7: & \quad \quad \textbf{If} $\delta<a$, \textbf{then} set $\mathcal{A}=\mathcal{G}$ and $\delta = a$.\\
    \;8: & \quad \quad Update $\mathcal{G}\longleftarrow\mathcal{G}\backslash\hat{\mathcal{W}}$\\
    \;9: & \quad \textbf{Until} $\mathcal{G}=\phi$ or $\mathcal{G}\in \mathbfcal{S}_1$\\
    \;10: & \textbf{End For}\\
    \;11: & Output $\mathcal{G}^*=\mathcal{A}, \varepsilon(\bH_{k},\mathcal{G}^*,\mathcal{S}\backslash\mathcal{G}^*,\mathbf{R})=\delta$ and stop.\\
\end{tabular}}}\\
\rule{\linewidth}{0.3mm}
\end{minipage}\vspace{.2 in}\\

In the following we consider the rate allocation for the group decoder.
Assume each user $k$ is allocated to a subset of antennas $\calS_{k} \subseteq \M$.
The rate $\bR = [R_{1}, R_{2}, ..., R_{M}]$ is decodable if for each receiver $k$ there exists a multi-stage decoding
defined by the partition $\{\mathcal{G}_{m}^k\}^{p_k+1}_{m=1}$,
where in stage $m$ the antennas in $\mathcal{G}_{m}^k$ with rates $[R_{t}]_{t \in \mathcal{G}_{m}^k}$
are decodable by treating $\bigcup^{p_k+1}_{t = m+1} \mathcal{G}_{t}^k$ as additive noise.
Given a target rate vector $\br = [r_{1}, r_{2}, ..., r_{M}]$, we aim to find a decodable rate vector $\bR = [R_{1}, R_{2}, ..., R_{M}]$
that maximizes the minimum rate increment $\min_{1 \leq t \leq M} (R_{t} - r_{t})$.

To this end, each receiver $k$ initializes the undecoded set $\D$ as $\M$,
and sequentially in each stage $m$ searches the group partition $\mathcal{G}^*$ such that,
\begin{equation}
\mathcal{G}^* = \argmax_{\mathcal{G} \subseteq \D} \varepsilon(\bH_{k},\mathcal{G},\D \setminus \mathcal{G}, \mathbf{\bR})
\end{equation}
and sets $\mathcal{G}_{m}^k = \mathcal{G}^*$ and updates the undecoded set $\D \leftarrow \D \setminus \mathcal{G}_{m}^k$ for $m = 1, 2, ..., p_k$,
until all its desired antennas $\calS_{k}$ are included in the decoded set $\bigcup^{p_k}_{m=1} \mathcal{G}_{m}^k$.
In each step $m$, receiver $k$ identifies the group partition $\mathcal{G}_{m}^k$,
and updates the rate for the antennas $t \in \mathcal{G}_{m}^k$ as follows,
\begin{equation}
R^k_{t} = r_{t} + \varepsilon(\bH_{k},\mathcal{G}_{m}^k,\D \setminus \mathcal{G}_{m}^k, \bR).
\end{equation}
The rates $R^k_t$ for $t \in \M \setminus \cup^{p_k}_{m=1} {\cal G}^k_m$ are set to be infinity since they are not required to be decoded.
The rate allocated to antenna $t$ is given by $R_{t} = \min_{1 \leq k \leq K} R^k_{t}$.
The detailed steps are given in Algorithm 3.

\begin{minipage}[h]{6.5 in}
\rule{\linewidth}{0.3mm}\vspace{-.1in}
{\bf {\footnotesize Algorithm 3 - Rate Allocation based on SGD}}\vspace{-.2in}\\
\rule{\linewidth}{0.2mm}
{ {\small
\begin{tabular}{ll}
    \;1: & Input $\br = [r_{1}, r_{2}, ..., r_{M}]$, and $\calS_{k}$\\
    \;2: & Initialize undecoded set $\mathcal{D} = \mathcal{M}$ and $m=1$\\
    \;3: & \textbf{for} $k = 1,2,...,K$ \\
    \;4: & \quad \textbf{Repeat}\\
    \;5: & \quad\quad \textbf{find} $\mathcal{G}^* = \argmax_{\mathcal{G} \subseteq \D} \varepsilon(\bH_{k},\mathcal{G},\D \setminus \mathcal{G}, \mathbf{\br})$\\
    \;6: & \quad\quad \textbf{update} $\mathcal{G}^k_m \leftarrow \mathcal{G}^*$ and $\mathcal{D} \leftarrow \mathcal{D} \setminus \mathcal{G}^k_m$\\
    \;7: & \quad\quad  \textbf{for} the users $t \in \mathcal{G}_{m}^k$, update $R^k_{t} = r_{t} + \varepsilon(\bH_{k},\mathcal{G}_{m}^k,\D \setminus \mathcal{G}_{m}^k, \br)$\\
    \;8: & \quad\quad \textbf{update} $m=m+1$ \\
    \;9: & \quad\textbf{Until} $\calS_{k} \subseteq \bigcup_{m=1}^{p_k}\mathcal{G}^k_m$\\
    \;10: & \textbf{End for} \\
    \;11: & Set $R^k_t = +\infty$ for all $t \in {\cal M} \setminus \bigcup^{p_k}_{m=1} {\cal G}^k_m$. \\
    \;12: & Output $R_{t} = \min_{1 \leq k \leq K} R^k_{t}$.\\
\end{tabular}}}\\
\rule{\linewidth}{0.3mm}
\end{minipage}\vspace{.2 in}\\

\subsection{Hybrid Group Decoding}\label{subsec.HybridDforVBS}

Note that the computational complexity of the group decoder is much higher than that of the MMSE decoder.
In order to reduce the decoding complexity, we propose a hybrid decoding scheme that switches to the MMSE decoder
when the MMSE decoder can decode the signal of interest.
More specifically, the MMSE filter at the $k$-th receiver is given by,
\begin{equation}\label{equ.MMSEFilter}
    \emph{\textbf{G}}_{k} = (\bH_{k}^\dag \bH_{k} + \sigma^2 \bI)^{-1}\bH_{k}^\dag.
\end{equation}
Let $\emph{\textbf{g}}_{k}^t$, $1 \leq t \leq M$, denotes the $t$-th row of $\emph{\textbf{G}}_{k}$.
The achievable rate from the $t$-th transmit antenna is given by
\begin{equation}\label{equ.MMSErates}
    \hat R_{t}^{k} =
    \log  \left(1+ \frac{|\emph{\bg}_{k}^t \emph{\bh}_{k}^{t}|^2}{|\emph{\bg}_{k}^t|^2 \sigma^2 + \sum_{q\neq t}^{M}|\emph{\bg}_{k}^q \emph{\bh}_{k}^{q}|^2} \right).
\end{equation}
The MMSE decoding is adopted at receiver $k$ if the achievable rate $\hat R_{t}$, $t \in \calS_{k}$, via MMSE decoding
can exceed the target rate $r_{t}$ by amount of $\delta$, i.e.,
\begin{equation}\label{equ.PracticeRates}
    \hat R_{t}^{k} - r_{t} \geq \delta, \ \ \mbox{for all} \ t \in {\cal S}_k .
\end{equation}

For the hybrid group decoder,
we first check whether each layer can be decoded by the MMSE decoding, i.e., whether \eqref{equ.PracticeRates} is satisfied.
If so, decode the received signal using the MMSE decoding,
otherwise decode the received signals using the SGD.
Note that only the rates based on the SGD are feedback to the transmitter.

We also employ channel codes with error detection capability,
which can detect the decoding error through the parity check of the decoded bits.
This can avoid the error propagation due to the decoding error in the signal cancelation.
In this work, when an error decoding happens for a signal layer, we do not perform the cancelation for that signal layer.

\section{Layer Extraction and Resource Block Allocation} \label{sec.UEPandSubA}

We aim to maximize the sum PSNR of all users via the layer extraction and resource allocation.
We adopt unequal error protection (UEP) scheme since the base layer should be more protected than the enhancement layers.
Moverover, we employ the auction algorithm for the RB allocation.

\subsection{User UEP Scheme with Channel Coding and Modulation}\label{subsec.MCS4UEP}
We employ an adaptive modulation and coding scheme (MCS) with QAM and finite-length practical channel codes.
We set the rate of transmit antenna $t$ of the $i^{th}$ RB as the achievable rate $R_{i,t}$ obtained from the rate allocation of OSGD,
assuming that all antenna signals are decodable at all receivers, i.e., ${\cal S}_k = {\cal M}$ for all receiver $k$.
This is to make sure that the allocated rates are decodable for all possible antenna allocations.

Considering the different quality of different video layers, we introduce the UEP,
which protects different video layers using code of different rates.
For the $l$-th quality layer of the $k$-th user, we define a coding-rate margin $\Gamma_k^l$,
In the MIMO-OFDM system, given the practical rate $R_{i,t}$ and the modulation constellation $S_{k}^l$,
if the $t$-th antenna of the $i$-th RB is allocated to the $l$-th layer of the $k$-th user,
the real transmission spectrum rate is given by
\begin{equation} \label{equ.UEP}
    \bar R_{i,t,k}^l = \left(R_{i,t}-\Gamma_k^l \log_2|S_{k}^l|\right)^+.
\end{equation}

In this work, the modulation schemes are selected from $\{QPSK, 16QAM, 64QAM\}$,
and the code rates are selected from $\C = \{1/4, 1/3, 1/2, 2/3, 3/4, 7/8\}$.
Note that for the M-QAM, the capacity of bit-interleaved coded modulation with Gray mapping well approximates that of the coded modulation,
and the capacity of coded modulation well approximates that of Gaussian modulation when the spectral efficiency is below $\frac 1 2 \log_2 M$ bits per channel use.
Therefore, for $\bar R_{i,t,k}^l \leq 1.0$ we associate it with QPSK, for $1.0 \leq \bar R_{i,t,k}^l \leq 2.0$
we associate it with 16QAM, and for $\bar R_{i,t,k}^l \geq 2.0$ we associate it with 64QAM.
Then, given the real transmission rate $\bar R_{i,t,k}^l$ and the associated modulation scheme $S_{i,t,k}^l$,
the real channel coding rate is given by
\begin{equation}
    \bar r_{i,t,k}^l =\frac{\bar R_{i,t,k}^l}{\log_2|S_{i,t,k}^l|}.
\end{equation}
In practice, the channel code rate is selected as the maximum rate $c_q \in \C$ smaller than or equal to $\bar r_{i,t,k}^l$, i.e.,
\begin{equation}\label{equ.CoderateMod}
    r_{i,t,k}^l =  \underset{\{c_q \in \C, c_q \leq \bar r_{i,t,k}^l \}}{\max} \; c_q.
\end{equation}

For practical channel codes, we employ LDPC codes due to its capacity approaching performance and the decoding error detection capability.

\subsection{Resource Allocation}\label{subsec.SubAllo}

In this work, to reduce the signalling overhead, we assign the same coding rate for the same layer,
even if it is transmitted across different RBs and/or different transmit antennas.
Our goal is to allocate the RBs and transmission antennas to the user video layers.

We define $a_{i,t,k}^l$ as an indicator of whether the $t$-th antenna of the $i$-th RB is allocated to the $l$-th layer of the $k$-th user.
Let ${\cal{A}}_k^l = \{(i,t)|a_{i,t,k}^l = 1\}$ be the set of the combinations of RBs and transmit antennas allocated to the $l$-th layer of the $k$-th user.
Then the transmission rate is given by $r^l_{k}\sum_{i=1}^{N_{rb}} \sum_{t=1}^M a_{i,t,k}^l$,
where $r^l_{k}$ is the rate allocated to the $l$-th layer of the $k$-th user.
Note that the real transmission rate of the $l$-th layer of the $k$-th user should be smaller than the transmission rate of all its allocated subcarriers, i.e.,
\begin{equation}\label{equ.RateCouple2}
r_{t}^l \leq \underset{(i,t) \in {\cal{A}}_k^l }{\min} r_{i,t,k}^l.
\end{equation}

Considering the video decoding dependency, for each user $k$,
the information extraction of the $l$-th layer is successful if and only if this layer
and all its lower layers are received correctly.
Hence the effective information extraction rate is given by
\begin{equation}\label{equ.RateCouple}
    C_k^l = r^l_{k}\sum_{i=1}^{N_{rb}} \sum_{t=1}^{M}a_{i,t,k}^l \left[\prod_{q=0}^{l-1} \mathbf{1}\left(\sum_{i=1}^{N_{rb}} \sum_{t=1}^{M}a_{i,t,k}^{q} r^{q}_{k}
    \geq V_{k,q} - V_{k,q-1}\right)\right],
\end{equation}
where $V_{k,q}$ is the total bit rate up to the $q$-th layer of the $k$-th user's reconstructed video,
the indicator denotes whether all information of each SVC layer can be transmitted.

Then the resource allocation problem is to maximize the average PSNR of all users,
which is $\bar{Q}= \frac{1}{K} \sum_{k=1}^{K}Q(r_k)$,
subject to the decoding dependency constraints.
Similar to \cite{ZanYang}, using \eqref{equ.RQModel}, we can formulate the resource allocation problem as follows,
\begin{subequations}\label{equ.ResourceAlloPr1}
\begin{alignat}{2}
    & \max_{a_{i,t,k}^l}
    & & \sum_{k=1}^{K} \sum_{l=0}^{L} \beta_k^l \min(C_k^l, V_{k,l} - V_{k,l-1}) \label{equ.ResourceAlloPr1sub0}\\
    & \mbox{s.t.} \quad
    & & \sum_{k=1}^{K} \sum_{l=0}^{L} a_{i,t,k}^l \leq 1, \; a_{i,t,k}^l \in \{0,1\}, \; \forall \;i, \; \forall \;t, \label{equ.ResourceAlloPr1sub1}\\
    &&& \sum_{i=1}^{N_{rb}} \sum_{t=1}^{M} a_{i,t,k}^0 r_{k}^0 \geq V_{k,0},\label{equ.ResourceAlloPr1sub2}\\
    &&& r_{t}^l \leq \underset{(i,t) \in {\cal{A}}_k^l }{\min} r_{i,t,k}^l, \; \forall \;t, \; \forall \;l.\label{equ.ResourceAlloPr1sub3}
\end{alignat}
\end{subequations}
The min$(\cdot)$ operator in \eqref{equ.ResourceAlloPr1sub0} means that
the extra allocation to the $l$-th layer exceeding its rate bound $V_{k,l} - V_{k,l-1}$ is a waste.
Constraint \eqref{equ.ResourceAlloPr1sub1} imposes that
each RB's transmission antenna can be assigned to at most one quality layer of one user;
constraint \eqref{equ.ResourceAlloPr1sub2} imposes that the base layer reconstruction quality must be satisfied for all users;
and constraint \eqref{equ.ResourceAlloPr1sub3} imposes that the transmission rate for each layer should be smaller than that of the allocated RBs.

Note that this resource allocation problem is of the same nature as the subcarrier allocation problem in \cite{ZanYang},
which can be solved using the low-complexity auction algorithm.

\section{Numerical Results and discussions} \label{sec.SimulationResults}

We use video sequences, Mobile and Soccer, encoded by the SVC using JSVM 8.5 \cite{JSVMsofrware}.
Both sequences are coded at fixed spatial (CIF, 352 $\times$ 288: 4:2:0) and temporal resolution (30 frames per second)
with medium grained scalability (MGS) for quality enhancement.
We consider the LTE transmission scenario.
More specifically, the time is split into frames,
each one composed of 10 consecutive transmission time intervals (TTIs),
each lasting for 1 ms.
In the frequency domain, the total bandwidth is divided into sub-channels of 180 kHz,
each one with 12 consecutive and equally spaced OFDM subcarriers.
For the transmission system under consideration,
we assume that the OFDM system has 72 subcarriers.
We assume independent channels for different users and the 3GPP Extended Pedestrian A (EPA) channel model is adopted for each user \cite{3GPPR36521}.
We employ the low-density parity-check (LDPC) code with $L_{ch}=5040$ symbols.
Set the threshold $\delta = 0.2$ and the maximum group size $\mu = 2$.
Assume that the modulation is selected from $\{QPSK, 16QAM, 64QAM\}$,
and the LDPC code rate is selected from $\C = \{1/4, 1/3, 1/2, 2/3, 3/4, 7/8\}$.

We make simulations for the multi-user MIMO-OFDM system shown in Fig.~\ref{Fig.Fig1MIMOdownlink}.
We obtain one base layer and 4 enhancement layers from the video encoder.
The first 298 frames (almost 10s) are used to assess the reconstructed video quality.
The UEP scheme is considered by setting target BLERs $P_e^0=0.001$, $P_e^1=0.01$,
and $P_e^2=0.1$ for the base layer, three temporal enhancement layers and one quality enhancement layer, respectively,
with the corresponding $\Gamma$ values $\Gamma^0_k = 0.15$, $\Gamma^1_k = 0.13$, and $\Gamma^2_k = 0.10$.
We compare the reconstructed video quality for both the MMSE decoder and the SGD-based hybrid decoder.
Note that we assume no transmission error for the header information.

\begin{figure}[htbp]
\center{ \epsfxsize=5.5in \centerline{\epsffile{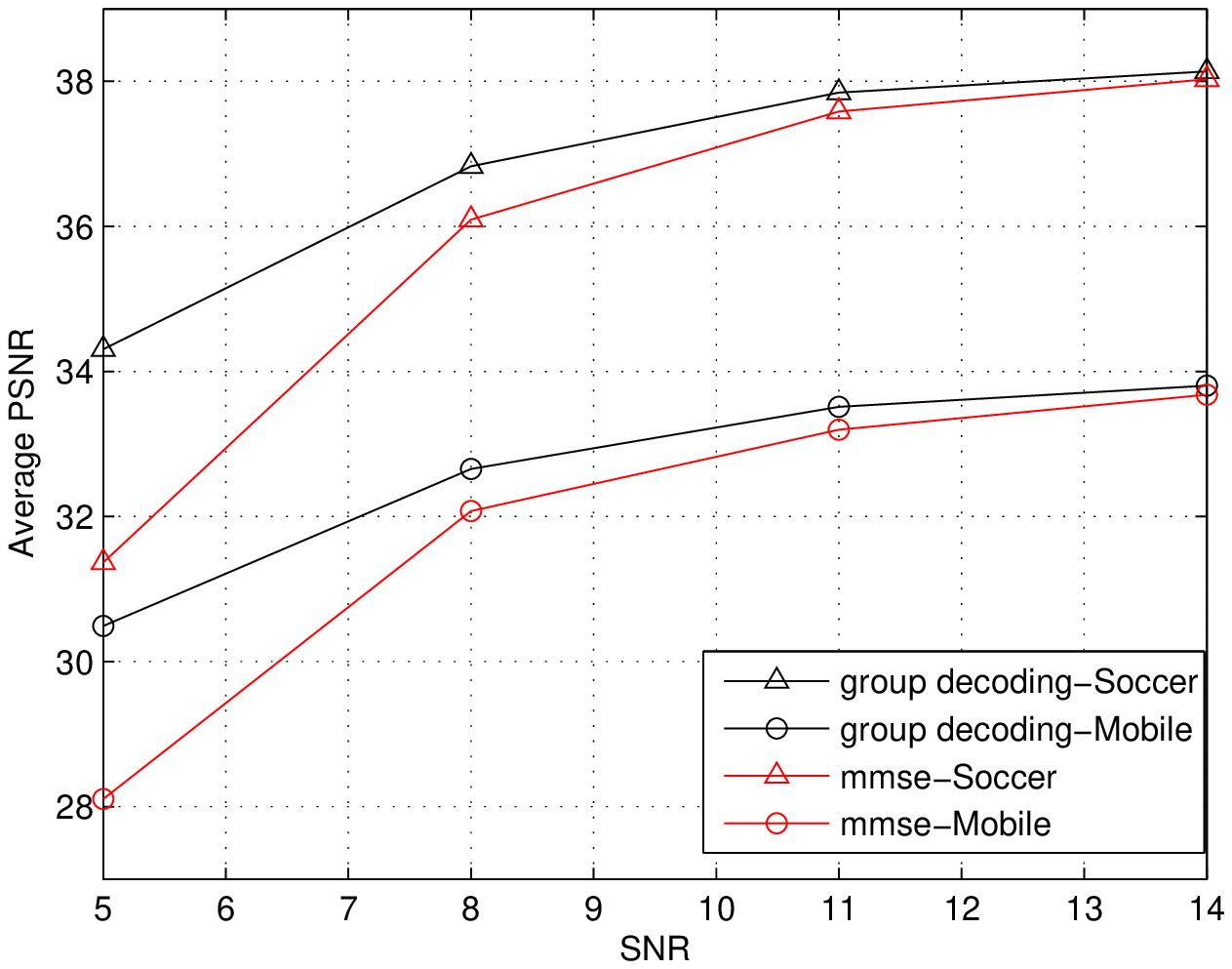}}
\caption{\small The average PSNR comparison between the SGD and MMSE decoding.}
\label{Fig.Fig3PSNRvsSNRS2}}
\end{figure}

We compare the reconstructed video quality of the SGD-based hybrid decoding
and the MMSE decoding, for test video sequences Soccer and Mobile.
The performances of the above two scenarios are shown in Fig.~\ref{Fig.Fig3PSNRvsSNRS2}.
In Fig.~\ref{Fig.Fig3PSNRvsSNRS2}, ``group decoding-Soccer'' and ``group decoding-Mobile'' denote the PSNR of the reconstructed video from Soccer and Mobile using the SGD-based hybrid decoding, respectively;
``mmse-Soccer" and ``mmse-Mobile" denote the PSNR of the reconstructed video from Soccer and Mobile with the MMSE decoding scheme, respectively.
We compare the average PSNR value of the first 298 frames for various channel SNRs.
It is seen that the reconstructed video sequence from the SGD-based hybrid decoding outperforms
that from the MMSE decoding by $0.12 \sim 2.7$dB.

\begin{figure}[htbp]
\center{ \epsfxsize=5.5in \centerline{\epsffile{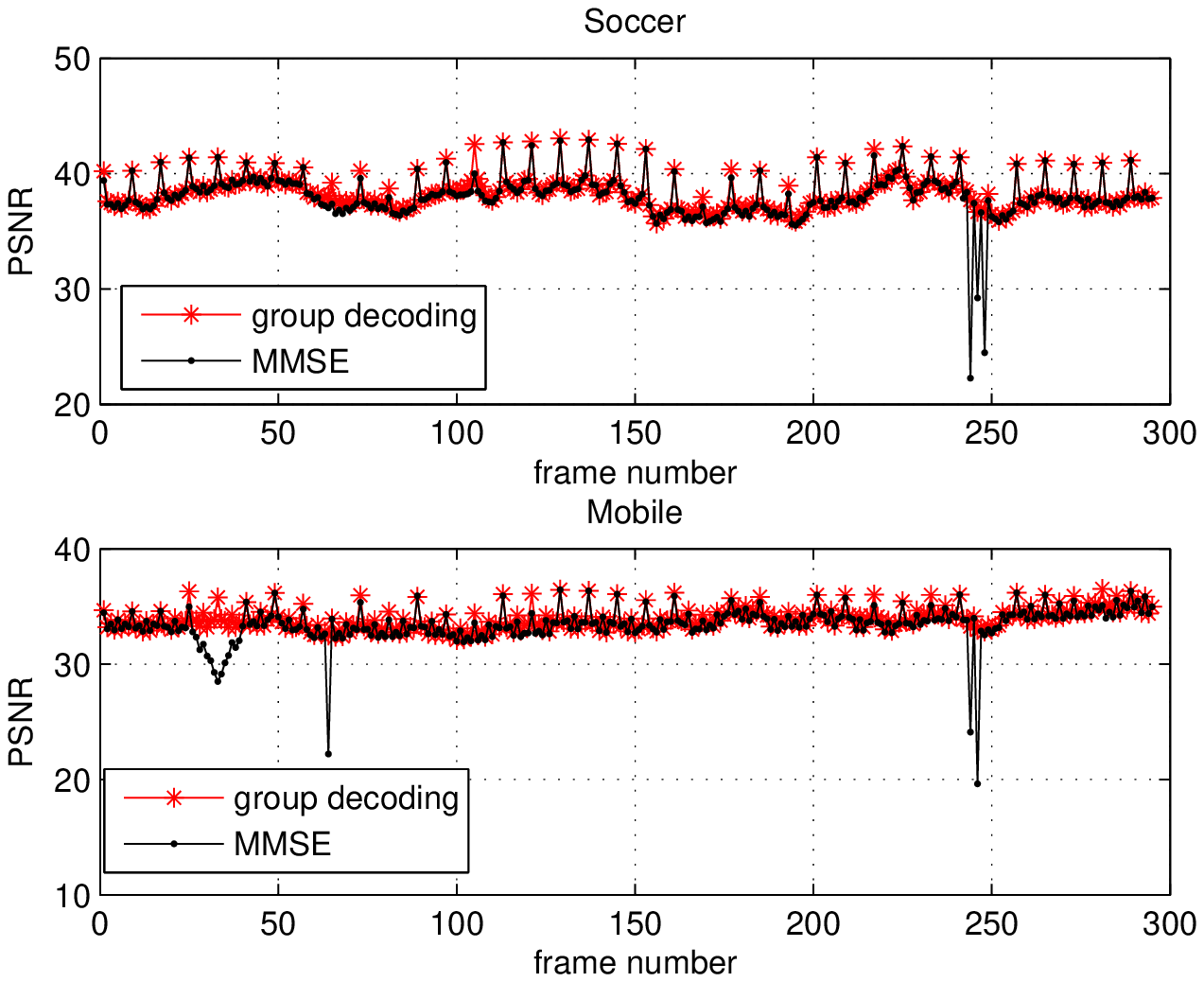}}
\caption{\small Per frame PSNR comparison between the SGD and the MMSE decoding.}
\label{Fig.Fig4soccerSNR}}
\end{figure}

In Fig.~\ref{Fig.Fig4soccerSNR}, we compare the per frame PSNR.
It is seen that the proposed SGD-based hybrid decoding exhibits PSNR improvement over the MMSE decoding scheme at frames 242-250 for ``Soccer'',
and frames 24-42, 63-65, and 242-248 for ``Mobile''.

\begin{figure}[htbp]
\center{ \epsfxsize=5.5in \centerline{\epsffile{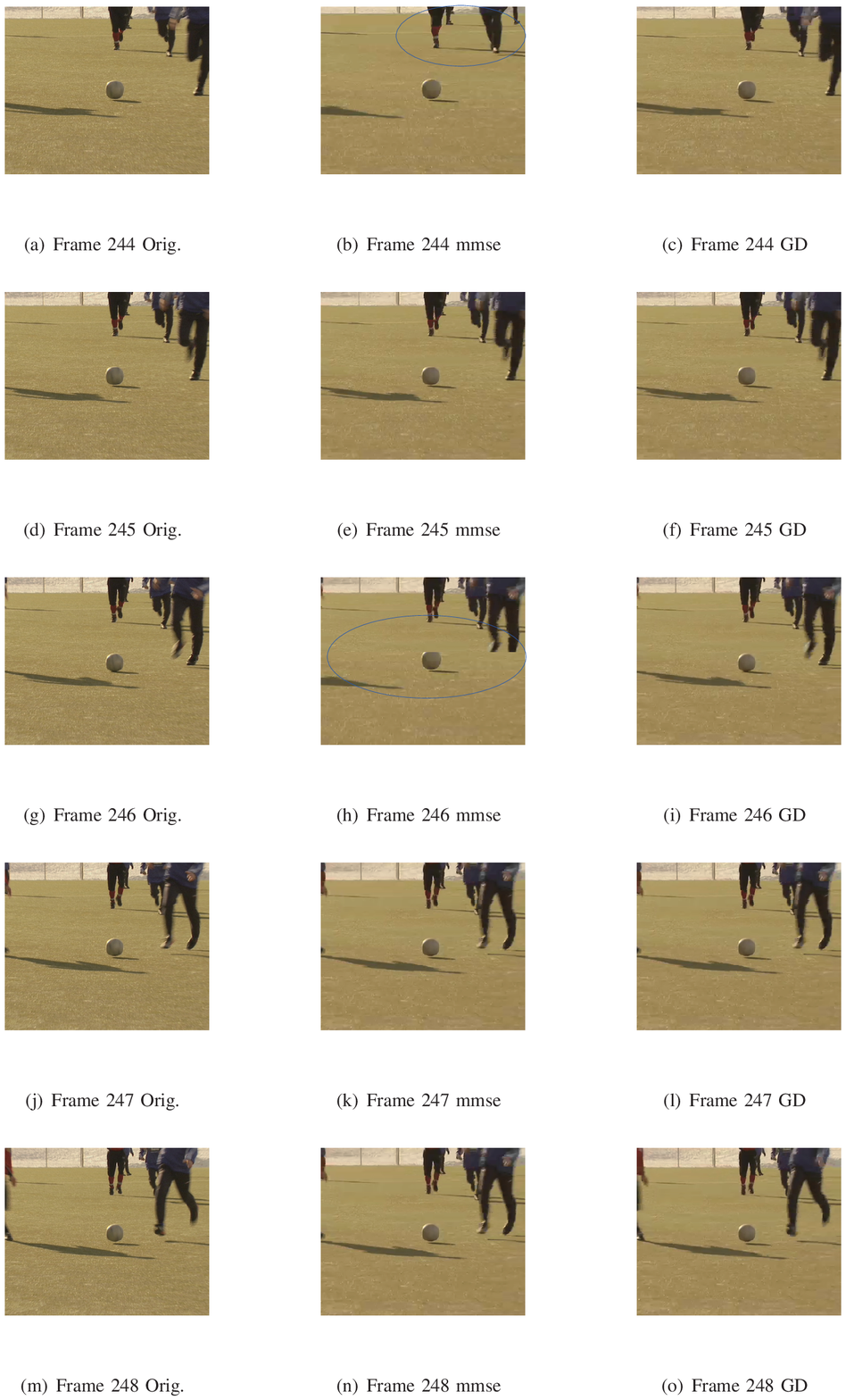}}
\caption{\small The example of different decoding schemes with Soccer.yuv: frame 244-248. ``Orig.'' denotes the original video frame; `mmse'' denotes the MMSE decoding scheme; and ``GD'' denotes the proposed hybrid decoding scheme.}
\label{Fig.Fig5Soccer5Frames}}
\end{figure}

In Fig.~\ref{Fig.Fig5Soccer5Frames}, we further compare the sampled video frames 244-248 for ``Soccer''.
The first column shows the original video sequence;
the second column shows the video sequence reconstructed by the MMSE decoding;
and the third column shows the video sequence reconstructed by the proposed SGD-based hybrid decoding.
There is almost no difference between the video sequences reconstructed by the proposed hybrid decoding and the original video sequences.
From sub-figures \ref{Fig.Fig5Soccer5Frames} $(b)$ and \ref{Fig.Fig5Soccer5Frames} $(h)$,
it is seen that the proposed SGD-based hybrid decoding scheme provides significant reconstructed video quality improvement compared with the MMSE decoding.

\section{Conclusions} \label{sec.Conclusion}
In this paper, we have proposed a scalable video broadcast scheme for downlink MIMO-OFDM systems
that employs successive group decoding (SGD).
Assuming different coding rate and modulation used for different video layers,
we have proposed a resource allocation that aims to maximize the sum PSNR of the reconstructed video sequences.
Simulation results have demonstrated that
the proposed scheme offers significant reconstructed video quality gain compared with the MMSE decoding.

\small{\baselineskip = 10pt
\bibliographystyle{./IEEEtran}
\bibliography{./shuying_2}
}

\end{document}

%% file: pream_bm.tex
\newtheorem{prop}{Proposition}

\newtheorem{cor}{Corollary}

\newtheorem{lm}{Lemma}

\newtheorem{thm}{Theorem}

\newcommand{\be}{\begin{eqnarray}}
\newcommand{\ee}{\end{eqnarray}}
\newcommand{\benn}{\begin{eqnarray*}}
\newcommand{\eenn}{\end{eqnarray*}}
\def\IR{\rm I \kern-0.20em R}
\newcommand{\utwi}[1]{\mbox{\boldmath $ #1$}}

\newcommand{\bthm}{\begin{thm}}
\newcommand{\ethm}{\end{thm}}

\newcommand{\bcor}{\begin{cor}}
\newcommand{\ecor}{\end{cor}}
\newcommand{\bprop}{\begin{prop}}
\newcommand{\eprop}{\end{prop}}
\newcommand{\blm}{\begin{lm}}
\newcommand{\elm}{\end{lm}}
\newcommand{\beq}{\begin{equation}}
\newcommand{\eeq}{\end{equation}}
\newcommand{\ber}{\begin{eqnarray}}
\newcommand{\eer}{\end{eqnarray}}

\newcommand{\bproof}{\begin{proof}}
\newcommand{\eproof}{\end{proof}}

\newcommand{\argmax}{\mathop{\mbox{\rm arg\,max}}}

\newcommand{\bit}{\begin{itemize}}
\newcommand{\eit}{\end{itemize}}
\newcommand{\ben}{\begin{enumerate}}
\newcommand{\een}{\end{enumerate}}
\newcommand{\bdesc}{\begin{description}}
\newcommand{\edesc}{\end{description}}
\newcommand{\beqarrn}{\begin{eqnarray*}}
\newcommand{\eeqarrn}{\end{eqnarray*}}
\newcommand{\bproofof}{\begin{proofof}}
\newcommand{\eproofof}{\end{proofof}}
\newenvironment{rem}{\begin{trivlist}\item[]{\bf
Remark:}\hspace{4mm}}{\end{trivlist}}
\newcommand{\brem}{\begin{rem}}
\newcommand{\erem}{\end{rem}}
\newenvironment{rems}{\begin{trivlist}\item[]{\bf
Remarks}\begin{itemize}}{\end{itemize}\end{trivlist}}
\newcommand{\brems}{\begin{rems}}
\newcommand{\erems}{\end{rems}}
\newtheorem{fact}{Fact}
\newcommand{\bfact}{\begin{fact}}
\newcommand{\efact}{\end{fact}}
\newtheorem{examp}{Example}
\newcommand{\bexamp}{\begin{examp}\rm}
\newcommand{\eexamp}{\end{examp}}
\newtheorem{defn}{Definition}
\newcommand{\bdefn}{\begin{defn}\rm}
\newcommand{\edefn}{\end{defn}}

\newtheorem{alg}{Algorithm}
\newcommand{\balg}{\begin{alg}}
\newcommand{\ealg}{\end{alg}}

\newtheorem{prob}{Problem}
\newcommand{\bprob}{\begin{prob}}
\newcommand{\eprob}{\end{prob}}

\newcommand{\bvtm}{\begin{verbatim}}
\newcommand{\bfig}{\begin{figure}}
\newcommand{\efig}{\end{figure}}
\newcommand{\bcen}{\begin{center}}
\newcommand{\ecen}{\end{center}}

\long\def\comment#1{}

\def \n2{{N_0 \over 2}}

\def \h5{\hspace{0.5in}}

\newcommand{\bg}{{\utwi{g}}}
\newcommand{\bh}{{\utwi{h}}}

\newcommand{\br}{{\utwi{r}}}

\newcommand{\bu}{{\utwi{u}}}

\newcommand{\bx}{{\utwi{x}}}
\newcommand{\by}{{\utwi{y}}}

\newcommand{\bH}{{\utwi{H}}}
\newcommand{\bI}{{\utwi{I}}}

\newcommand{\bR}{{\utwi{R}}}

\newcommand{\dff}{\stackrel{\triangle}{=}}